# Hybridization-controlled charge transfer and induced magnetism at correlated oxide interfaces

M.N. Grisolia[1], J. Varignon[1♦], G. Sanchez-Santolino[2♦], A. Arora[3], S. Valencia[3], M. Varela[4,2], R. Abrudan[3,5], E. Weschke[3], E. Schierle[3], J.E. Rault[6], J.-P. Rueff[6], A. Barthélémy[1], J. Santamaria[2] and M. Bibes[1]*

[1] Unité Mixte de Physique CNRS/Thales, 1 avenue A. Fresnel, 91767 Palaiseau, France, and Université Paris-Sud, 91405 Orsay, France.

[2] GFMC, Departamento Física Aplicada III, Universidad Complutense Madrid, 28040 Madrid, Spain, and Laboratorio de Heteroestructuras con aplicación en Spintronica, Unidad Asociada CSIC/Universidad Complutense de Madrid, Sor Juana Inés de la Cruz, 3, 28049 Madrid, Spain.

[3] Helmholtz-Zentrum Berlin für Materialen & Energie, Albert-Einstein-Strasse 15, 12489 Berlin, Germany.

[4] Materials Science & Technology Division, Oak Ridge National Laboratory, Oak Ridge, TN 37831, USA.

[5] Institut für Experimentalphysik/Festkörperphysik, Ruhr-Universität Bochum, 44780 Bochum, Germany.

[6] Synchrotron SOLEIL, L'Orme des Merisiers Saint-Aubin, BP 48, 91192 Gif-sur-Yvette, France

At interfaces between conventional materials, band bending and alignment are classically controlled by differences in electrochemical potential. Applying this concept to oxides in which interfaces can be polar and cations may adopt a mixed valence has led to the discovery of novel two-dimensional states between simple band insulators such as $LaAlO_3$ and $SrTiO_3$. However, many oxides have a more complex electronic structure, with charge, orbital and/or spin orders arising from correlations between transition metal and oxygen ions. Strong correlations thus offer a rich playground to engineer functional interfaces but their compatibility with the classical band alignment picture remains an open question. Here we show that beyond differences in electron affinities and polar effects, a key parameter determining charge transfer at correlated oxide interfaces is the energy required to alter the covalence of the metal-oxygen bond. Using the perovskite nickelate ($RNiO_3$) family as a template, we probe charge reconstruction at interfaces with gadolinium titanate $GdTiO_3$. X-ray absorption spectroscopy shows that the charge transfer is thwarted by hybridization effects tuned by the rare-earth (R) size. Charge transfer results in an induced ferromagnetic-like state in the nickelate, exemplifying the potential of correlated interfaces to design novel phases. Further, our work clarifies strategies to engineer two-dimensional systems through the control of both doping and covalence.

* : manuel.bibes@thalesgroup.com ; ♦ : these authors contributed equally to this work

In transition metal perovskites $ABO_3$ strong Coulomb repulsion (correlations) influences dramatically the energy and occupancy of into $t_{2g}$ (triplet) and $e_g$ (doublet) states. This has a dramatic influence on the physical properties: for instance, correlations can yield an insulating state, even when electrons partially fill the *d* bands (in contrast with classical band insulators that have no *d* electrons)[1,2]. Correlations induce a local interdependence of the charge, spin and orbital degrees of freedom from which magnetic order and/or orbital order can emerge. The flexibility of the perovskite structure makes it possible to tune the electronic, magnetic and orbital states by varying the ionic radii of the A site cations, which modifies the bond lengths, angle and distances, thus controlling the electron band width W.

In addition, metal 3*d* and oxygen 2*p* bands (separated by an energy $\Delta$, known as the charge transfer energy) tend to hybridize[2–5]. As a result, electronic states usually have a mixed *p-d* character[6]. For early transition metals such as Ti and V, the oxygen 2*p* band lies well below the 3*d* band. Thus, hybridization is weak (purely ionic limit) and the conduction band has a preferred 3*d* character (*Mott-Hubbard* insulator). For later transition metal oxides such as rare earth $RNiO_3$ nickelates[7] (with $Ni^{3+}$ formally in a $3d^7$ configuration), the 2*p* band lies very close in energy to the 3*d* band (small $\Delta$, *charge transfer* insulator), or even above it ($\Delta$<0, *negative charge transfer* insulator). Hybridization between the oxygen 2*p* band to the 3*d* band is then very strong and for nickelates electronic states must be described as a superposition of the form $|\Psi\rangle = \alpha|3d^7\rangle + \beta|3d^8\underline{L}\rangle$ where $\underline{L}$ stands for the ligand (oxygen) holes ($\alpha^2 + \beta^2 = 1$). The system is now partly covalent, and the degree of covalence is given by the ratio between $\beta^2$ and $\alpha^2$. In the strongly covalent limit ($\beta^2/\alpha^2 >> 1$), conduction has a dominant *p* character (in contrast, in ionic Mott-Hubbard insulators $\beta^2/\alpha^2 << 1$) [2,8–11].

Small or negative charge transfer is a key property of compounds that, when doped, produce a superconducting state[12]. This finding has prompted proposals of confinement- and strain-engineered multilayers based on nickelates whose electronic structure would mimic those of cuprate parent materials[13,14]. Controlling doping in these systems is essential to achieve bound states between oxygen holes and *d* electrons (Zhang-Rice singlets[15,16]), believed to play a crucial role in high-$T_C$ superconductivity[3]. To that end, charge transfer at epitaxial oxide interfaces aims at exploiting the interplay between doping and correlated electron physics to nucleate novel electronic states with engineered correlations[17].

In this work, we study heterostructures combining rare-earth nickelates and $GdTiO_3$ (a ferromagnetic Mott-Hubbard insulator, GTO) because of the strong difference in electron affinities ($\Delta\phi$=1.2 eV, Ref.[18]), from which a large electron transfer from GTO to the nickelate is expected[17,19]. To probe the influence of covalence in the nickelate on this doping process, we change the rare-earth from La to Nd to Sm which will decrease the bond angle between the nickel and the oxygen ion, thus leading to less hybridization between the *p* and *d* bands. To investigate interfacial doping into perovskite nickelate thin films, we combine first-principles calculations with hard X-ray photoemission (HXPS) and X-ray absorption spectroscopy (XAS). In particular we examine details of Ni $L_{2,3}$, Ti $L_{2,3}$ and O *K* absorption edges which are known to provide direct information on charge and hybridization states. Further, we use X-ray magnetic circular dichroism (XMCD) and X-ray resonant magnetic scattering (XRMS) to probe magnetism in both GTO and the nickelates.

We have grown nickelate ($LaNiO_3$ - LNO, $NdNiO_3$ - NNO, $SmNiO_3$ - SNO) films on top of GTO films using pulsed laser deposition using $LaAlO_3$ (LAO) substrates (*cf*. Methods and Ref. [20,21]). Fig. 1a displays reflection high-energy electron diffraction (RHEED) patterns before growth (LAO substrate, left), after growth of the GTO film (centre) and after deposition of the LNO film (right), indicative of two-dimensional growth. Clear RHEED oscillations were observed while growing the GTO film, which we used to fix the GTO thickness to 7 u.c. in all samples. For the nickelates, no oscillations were observed but a large RHEED intensity was recovered during deposition. Through grow rate calibrations by X-ray reflectometry, the nominal thickness of the nickelate layers was set to 7 u.c. as well. Azimuthal RHEED analysis along with X-ray diffraction experiments attested of the epitaxial character of all three bilayers (LNO/GTO, NNO/GTO and SNO/GTO).

Fig. 1b shows a high resolution Z-contrast scanning transmission electron microscopy (STEM) image of a LNO/GTO sample, confirming epitaxy and sample quality. Low magnification images and compositional EELS maps obtained from the analysis of the Ti $L_{2,3}$, La $M_{4,5}$, Ni $L_{2,3}$ and the Gd $M_{4,5}$ signals reveal layers that are flat and continuous over long lateral distances. Occasional defects are observed in high magnification images (Fig. 1c) but the samples are epitaxial and sample quality is high. Elemental maps with atomic resolution reveal sharp interfaces between the LAO substrate and the GTO film, and between the GTO film and the nickelate layer. As expected, the LNO and GTO thicknesses were equivalent, within about one unit-cell.

We have first studied possible electronic reconstructions between a rare-earth titanate and a rare-earth nickelate using first-principles calculations. We used standard density functional theory (DFT) methods plus a Hubbard *U* correction chosen to reproduce correctly the bulk properties of both titanates and nickelates (see SI and Methods). We performed geometry relaxations on a $(GdTiO_3)_7/(GdNiO_3)_7$ superlattice strained on a LAO substrate. The computed average valence states per $BO_2$ plane of Ti and Ni atoms along the growth direction are summarized in Fig. 2a. Away from the interface between GTO and the nickelate, the Ti valence is close to 3+ as in the bulk but closer to the interface, the Ti valence increases from 3+ to 4+. An opposite behaviour is observed on the nickelate side: the nickel valence decreases from 3+ towards 2+ upon approaching the interface. This clearly indicated charge transfer, which spreads over 2-3 u.c. with a main contribution coming from the interfacial plane.

To look for changes in the valence of Ti ions at the interface, we have measured HXPS of Ti $2p_{3/2}$ core-levels on a LNO/GTO sample at two different photon incidence angles to probe either mainly the very first planes in the GTO layer (0.2 deg, grazing incidence) or the whole bilayer (10 deg), see the cyan and blue dots in Fig. 2d. Both spectra clearly consist of multiple components. Those peaked at low binding energies are strongly enhanced in grazing incidence, which we interpret as reflecting the presence of a built-in electric field shifting the core levels of interfacial Ti ions to higher energies (see also SI). To fit the experimental spectra we assume that the signal from each Ti plane has a Gaussian line shape, progressively shifted in binding energies, assuming a potential drop of 1.6 eV over ~3 u.c. in the GTO close to the interface (consistent with the 1.69 eV drop predicted from calculations). Fig. 2b shows the dependence of the energy shift as a function of the position in the GTO film. In a rigid band model, this upward band bending results from a progressive depletion of the 3*d* band as one gets closer to the interface *i.e.* it corresponds to an increase of the Ti valence from 3+ towards 4+ near the interface.

We have confirmed the change of the Ti valence near the GTO/LNO interface using EELS at the O *K*-edge. Indeed, the intensity of the prepeak in the O *K*-edge spectrum is proportional to the number of holes in the hybridized *p*-*d* band. In titanates, in which the conduction band is more *d*-like it is directly related to the number of holes in the *d* band, set by the oxidation state[22] *i.e.* the peak is enhanced for $Ti^{4+}$ compared to $Ti^{3+}$. Fig. 2e shows O K-edge spectra collected at different distances from the interface in the GTO layer. The peak area appears to increase as one approaches the interface (from darker to lighter blue). The data are plotted as solid squares in Fig. 2c, along with the normalized height of the O K-edge prepeak (blue open dots). Both sets of data indicate an increase of the prepeak intensity near the interface with the nickelate, consistent with a Ti valence increasing towards 4+. Interestingly, a "bulk-like" response is also recovered beyond 2 to 3 GTO planes. Both HXPS and EELS data thus point to an increase of the Ti valence towards 4+ occurring over 2 to 3 unit-planes near the interface with the nickelate. These experimental results are thus in qualitative agreement with the theory results of Fig. 2a.

To detect possible changes in the Ni valence near the interface, we conducted valence band HXPS of a LNO/GTO bilayer at 0.2 and 10 deg incidence angles. Here, the relative contribution of interfacial Ni planes to the overall signal is slightly higher at 10 deg than at 0.2 deg. In the spectra shown in Fig. 2g, two peaks (labelled "a" and "b") are visible at ~1.4 eV and ~0.2 eV. In single LNO films, they are ascribed to $t_{2g}$ and $e_g$ peaks, respectively. Their intensity is weaker than in thicker LNO films consistent with earlier results on ultrathin films[23]. Whereas peak a is larger at 10 deg, peak b is more intense in grazing incidence (see also the intensity difference plotted in Fig. 2h). This is surprising because in grazing incidence, both peaks are usually depressed in LNO single films (possibly reflecting a reduced metallicity at the film surface, *cf.* SI). Here, the stronger reduction of the intensity of peak b in the 10 deg spectra (for which the interface region contributes much more to the signal than at 0.2 deg) rather suggests a transfer of spectral weight to lower energies for interfacial planes. It is hard to extract quantitative information from these data, but for illustrative purposes we represent in Fig. 2f the intensity ratio of peaks a and b and the region probed at each incidence angle, and sketch the possible associated trend in the spectral weight transfer across the LNO layer. The increase of spectral weight at low energy near the interface is consistent with a local increase in the number of electrons in the nickelate, in line with first-principles calculations. In summary, both DFT calculations and experimental data indicate the transfer of charge from the titanate to the nickelate across the interface.

We now move on to study the influence of the rare earth in the nickelate on the interfacial charge transfer. We have used XAS at the Ti and Ni $L_{2,3}$ edges in LNO/GTO, NNO/GTO and SNO/GTO bilayers and different reference samples. In Fig 3a, we plot the XAS at the Ni $L_{2,3}$ edge for the three bilayers, corrected for the overlap of the La $M_{4,5}$ signal (see SI). In Ni oxides, the spectral shape of the Ni $L_2$ edge strongly varies with the Ni oxidation state[8,24] and we thus use it to gain insight into the Ni valence in our samples. As visible in Fig. 3b, the shape of the spectra in our bilayers clearly deviates from that of a reference LNO film (top panel), which readily indicates a different formal Ni oxidation state. The enhancement of the low energy peak (labelled "a") is reminiscent of the signature of $Ni^{2+}$ in the NiO reference[24] (bottom panel). We analyse the valence change in the different bilayers by comparing the relative intensity of the two peaks ("a" and "b"). Fig. 3c summarizes these results and suggests that the Ni valence is reduced towards 2+ for the three samples, with a stronger change for the NNO/GTO and SNO/GTO samples. This behaviour is further confirmed by our DFT calculations on a set of different RTO/RNO superlattices (Fig. 3d and SI) revealing a strong effect of the sole rare earth in the nickelate on the amount of transferred electrons.

In parallel, we looked for variations in the Ti valence by analyzing the XAS at the Ti $L_{3,2}$ edge, cf. Fig. 3e. Literature on XAS in rare-earth titanates is scarce[25] especially for strongly distorted compounds such as GTO, and direct comparison with bulk spectra is not possible. Trends in the XAS spectra can however be inferred from the electronic structure of perovskite titanates[26]. There are several important differences in the Ti 3*d* levels of a $Ti^{3+}$ perovskite compared to those of a $Ti^{4+}$ perovskite, including the finite *d* level occupancy and the splitting of both $t_{2g}$ and $e_g$ levels due to Jahn-Teller and orthorhombic $GdFeO_3$ distortions for $Ti^{3+}$ (Ref. [26]). As a result, the spectral shape and width of the absorption peaks are expected to be broader in rare-earth titanates than in $Ti^{4+}$ perovskites such as $SrTiO_3$ (STO), *cf.* Ref. [25,27]. In Fig. 3f, we plot the high-energy Ti $L_3$ peak for LNO/GTO, NNO/GTO and SNO/GTO bilayers, along with those of a GTO single film and a STO crystal. While the peak for STO can be well fitted by a single component, two components are needed for the GTO film and the three bilayers. The total width of the experimental peak as well as the energy splitting between both components decreases upon going from the GTO film through the bilayers series (see Fig. 3g). Because the structural parameters of GTO in the different bilayers are the same, this variation is unlikely to be associated with changes in the crystal field splitting. Rather, this finding suggests a mixed $Ti^{3+/4+}$ character for the bilayers, with an increase of the Ti valence towards 4+ for smaller rare-earths. Again, this behaviour is reproduced by our first-principles calculations (Fig. 3h). Remarkably, the trend is opposite and complementary to that of the Ni valence inferred from Fig. 3a-h, strongly suggesting that the transfer of electrons from the titanate to the nickelate is controlled by the sole rare-earth in the nickelate, in agreement with the calculations (*cf*. SI). Because here XAS probes the whole nickelate and titanate thickness, the valence changes that we deduce correspond to weighted averages within each layer. In line with recent results on $LaTiO_3/LaFeO_3$ interfaces[28] the total charge transferred from the GTO to the nickelate is thus large, particularly for the NNO/GTO and SNO/GTO samples. For LNO/GTO however, charge transfer is thwarted by strong covalence effects as we shall see in the following.

We now turn to investigate how the transferred charge is distributed in the hybridized bands of the nickelate. As shown by Van Veenendaal *et al*[29], changes in the multiplet splitting of the Ni $L_3$ absorption edge can be used to estimate the level of covalence. This splitting corresponds to the energy separation between $t_{2g}$ and $e_g$ levels resulting from the interplay between hybridization and Coulomb repulsion that are both stronger for $e_g$ levels[6]. Previous studies in bulk samples have shown that progressing in the nickelate series (decreasing the rare earth ionic radius) results in a larger splitting of the $L_3$ peaks, analyzed in terms of a decrease in covalence[29,30]. We have analyzed and fitted the XAS at the Ni $L_3$ edge for LNO, NNO and SNO reference films as well as for LNO/GTO, NNO/GTO and SNO/GTO bilayers. Results for LNO and LNO/GTO, as well as for SNO and SNO/GTO are shown in Fig. 4a and 4b, respectively. The energy splittings between both components of the fit are plotted in Fig. 4c. The peak splitting increases upon going from LNO/GTO to NNO/GTO to SNO/GTO, reflecting a decrease in covalence, as found in bulk nickelates with decreasing rare-earth size[29]. In addition, we also find that the splitting values are similar between the bilayers and reference single films (fully strained on LAO). This indicates that the relative weights (probability) in the hybridized states $|\Psi\rangle = \alpha|3d^7\rangle + \beta\,|3d^8\underline{L}\rangle$ are comparable in films and in bilayers, suggesting that for a given rare-earth nickelate, the proportion $\beta^2/(\beta^2+\alpha^2)$ is characteristic of the chemical $Ni^{3+}$ state. To reinforce this interpretation, we performed simulations of the electronic structure of single films of nickelates (bulk and strained on LAO). The evolution of the computed covalence (estimated from the number of oxygen *p* electrons $N_p$ and the number *d* electrons $N_d$ near the Fermi level, *cf.* SI) as a function of the rare-earth nickelate presented in Fig. 4d is in close agreement with what is observed experimentally. This confirms the idea that covalence is mostly a local and intrinsic property of the Ni-O-Ni bond (with strain playing a minor role in the low strain limit of larger rare-earth nickelates on LAO).

In Fig. 4e, we plot the spectra for the three bilayers and two reference samples (purple: SNO single film, blue: GTO single film). Bilayers display two prepeaks at 529 eV and 531 eV, associated with oxygen hybridization with Ni and Ti, respectively. Upon going from LNO/GTO to NNO/GTO and SNO/GTO, the Ni prepeak intensity globally decreases while that of the Ti prepeak increases (see Fig. 4f). This latter trend is a direct confirmation of the change in the Ti valence discussed in Fig. 3c and 3g. The decrease in the Ni prepeak intensity as the rare earth is varied from La to Sm is much stronger than that found in bulk $RNiO_3$ (Ref. [31]) and thus cannot be attributed to a simple decrease in the covalence.

To understand this variation, it is helpful to consider that charge transfer from GTO causes a change from $Ni^{3+} = \alpha|3d^7\rangle + \beta|3d^8\underline{L}\rangle$ towards a $Ni^{2+} \approx \alpha'|3d^8\rangle$ (which is mostly a non-covalent state due to the very high correlation energy cost of states with $3d^9$ components). In the bilayers nickel can thus be described as a mixture of $Ni^{3+}$ and $Ni^{2+}$, *i.e.* $\alpha|3d^7\rangle + \beta|3d^8L\rangle + \alpha'|3d^8\rangle$, with now $\alpha^2+\beta^2+\alpha'^2=1$. Charge transfer from GTO increases $\alpha'^2$, thereby reducing $\alpha^2+\beta^2$ and in fact *both* $\alpha^2$ and $\beta^2$ since, as discussed above, the ratio $\beta^2/(\alpha^2+\beta^2)$ is fixed for a given nickelate (*cf.* Fig. 3c). This *rehybridization* process[32] requires a redistribution of electrons between the $|3d^7\rangle$ and the $|3d^8\underline{L}\rangle$ states, which occurs at an energy cost $\Delta$ for each electron transferred from $|3d^8\underline{L}\rangle$ to $|3d^7\rangle$. Their relative proportion is $\beta^2$ and thus the resulting energy cost is proportional to $\beta^2\Delta$.

Fig. 4g also presents the full width at half maximum (FWHM) of both prepeaks[33], characterizing the hole bandwidths for Ti – that increases as the Ti 3*d* band is depleted – and for Ni – that decreases as the Ni mixed states are filled. This observation definitely confirms the transfer of charge from the GTO to the nickelate, controlled by the rare earth size.

We summarize these ideas in Fig. 4h-j. The top panels correspond to a schematic representation of the electronic structure of "bulk" GTO (left) and nickelate (decomposed into $\alpha|3d^7\rangle + \beta|3d^8L\rangle$ states, right). The rare earth size decreases from (h) to (i) to (j), causing a progressive decrease of the bandwidth of the $3d^7$ state and a reduction of the covalent character (decrease of $\beta^2$). When an interface is built between GTO and the nickelate (bottom panels), the system attempts to transfer electrons in order to align the electrochemical potentials and this leads to the rehybridization process discussed above. For large rare-earth size (Fig. 4h), the nickelate has a strong covalent character, which increases the cost of rehybridization. This mechanism competes with the energy gain associated with the difference in electron affinity, thereby limiting the amount of charge that can be transferred across the interface. This scenario can also be viewed as a decrease of the effective $\delta\phi$ for larger rare earth radii.

We now turn to the magnetic properties of the heterostructures. Magnetism in transition metal insulators such as nickelates and titanates is ruled by super-exchange interactions[34]. Strongly distorted rare-earth titanates such as GTO are ferromagnetic[26]. Nickelates are antiferromagnets at low temperatures although ferromagnetic and antiferromagnetic exchange paths alternate in the [111] direction in a way determined by the ionic (antiferromagnetic) or covalent nature (ferromagnetic) of the bonds[7]. At the interface between GTO and a nickelate there is a $Ti^{3+}$-O-Ni antiferromagnetic super-exchange path mediated by $t_{2g}$ electrons which will locally induce a ferromagnetic-like moment in the nickelate. This path will be interrupted for the non-magnetic $Ti^{4+}$ if charge transfer occurs at the interface. Thus, looking at the magnetic interactions in the nickelate provides further information about covalence and charge transfer.

In Fig. 5, we present XAS, XMRS and XMCD measured at low temperature and 5 T in a NNO/GTO bilayer. A magnetic signal is detected at the Gd $M_{4,5}$ (we estimate $M_{Gd}$=5.8 $\mu_B$) and Ti $L_{2,3}$ edges, consistent with the ferromagnetic character of GTO. Interestingly, although bulk NNO is antiferromagnetic, a sizeable dichroic signal is measured at the Ni $L_{2,3}$ edge (sum rules leads to a value of about $M_{Ni}$=0.1-0.2 $\mu_B$ averaged over the whole 7 u.c. thickness).

The magnetic field dependence of the dichroic response at the Ni $L_2$ edge for this sample is shown in Fig. 6a-h. Panels Fig. 6a to Fig. 6f show XRMS difference spectra measured at different fields ramping the field down from 3T (darker graphs) and up from -3T (lighter graphs). A first observation is that at high magnetic field the spectra for both sweep directions virtually superimpose (Fig. 6a and 6f), while they are different at low magnetic field (see for instance Fig. 6d and 6e). This indicates hysteresis and a ferromagnetic-like behaviour. Secondly, the shape of the signal changes with the magnetic field. While this could reflect a magnetic-field induced change of the electronic and magnetic structure, such as a spin-state transition, the most plausible explanation is that several magnetic components are present in the sample, and that they respond differently to the magnetic field. Indeed, Ni is in a mixed 2+/3+ valence state and $Ni^{2+}$ and $Ni^{3+}$ may show different magnetic behaviour. Even with multiplet calculations, decomposing the experimental spectrum at each magnetic field into two dichroic signals unambiguously ascribable to $Ni^{2+}$ and to $Ni^{3+}$ would however be very delicate (notably due to the depth dependence of both Ni species). Here we simply fit the spectra with two lorentzian components for which we allow the width and position to vary only in a narrow range (the decomposition is given in the SI).

The amplitude of the first peak ("a", centred near 870 eV) is always positive and shows little magnetic field dependence even at high fields (*cf.* inset of Fig. 6g), as already observed in some magnetic multilayers[35,36]. The second peak ("b", see Fig. 6g) shows a much stronger dependence of its amplitude with magnetic field, corresponding to an hysteresis loop that is open at low positive fields, and shifted both horizontally and vertically. Such horizontal and a vertical loop shifts[37] – sometimes persisting up to several teslas[35,36] – are reminiscent of the phenomenon of exchange bias[38] which occurs at interfaces between a ferromagnet and an antiferromagnet[39]. We note that exchange bias has been observed in nickelate/manganite multilayers[40].

To explore the influence of interfacial charge transfer on magnetism in the nickelate, we compare XMCD at the Ni $L_{2,3}$ edge for the three bilayers after normalization at the $L_3$ peak of the respective XAS spectra, *cf*. Fig. 6i-k. As visible in Fig. 6i, a large XMCD signal is observed for all samples. In Fig. 6j, we highlight the signal at the Ni $L_2$ edge. As in the XAS (see Fig. 3b), the spectral shape varies with the rare-earth, suggesting an influence of the Ni valence change on its magnetic response. The intensity of the low energy peak "a" increases when going from LNO/GTO to SNO/GTO, as summarized in Fig. 6k. The parallel between Fig. 3c and Fig. 6k indicates that $Ni^{2+}$ ions strongly contribute to the overall magnetic signal, which provides further evidence of the role played by covalence in this system. Further work is needed to ascertain the exact nature of the exchange paths at play at this complex interface, but one may speculate that covalent exchange, proposed as the source of ferromagnetism in $Li_{1-x}Ni_{1+x}O_2$ (Ref. [41])¸ is an important ingredient to produce the observed Ni dichroism.

In summary, we have discovered that at interfaces between strongly correlated oxides, the covalent character of the transition metal / oxygen bonds plays a key role in determining the amount of charge transferred across the interface, and the resulting charge density in the different available bands. The situation is thus very different from that found at weakly correlated interfaces[42] where, conventionally, differences in electron affinity and possible polar discontinuities control the interfacial doping. Our results suggest novel strategies to engineer two-dimensional states in correlated oxides[43] by playing on both electron affinity mismatch and the level of covalence. In bulk nickelates, the latter can be adjusted by changing the rare earth size, but any mechanism altering the electron bandwidth should play the same role. For instance, strain-engineering in the large strain limit seems quite promising as an additional knob to craft interfacial properties. Another means of tuning covalence could be to move from oxygen to other ligands such as S or Se. All combined, these different handles may allow to finely tune the orbital hierarchy as well as the *p* and *d* hole density, in the search for artificial cuprate-like superconductivity.

### Acknowledgements

The authors thank Masashi Watanabe for the Digital Micrograph PCA plug-in, Flavio Y. Bruno for his assistance at the early stage of this project and Vincent Garcia and Richard Mattana for useful comments. The research leading to these results has received funding from the European Community's Seventh Framework Programme (FP7/2007-2013) under grant agreement #312284. Research at CNRS/Thales was supported by the ERC Consolidator Grant #615759 "MINT" and the region Île-de-France DIM "Oxymore" (project NEIMO). Research at ORNL was supported by the US Department of Energy, Office of Science, Basic Energy Sciences, Materials Sciences and Engineering Division. Work at UCM supported by grants MAT2014-52405-C02-01 and Consolider Ingenio 2010 - CSD2009-00013 (Imagine), by CAM through grant CAM S2013/MIT-2740 and by the ERC Starting Investigator Grant #239739 STEMOX. J.S. thanks the Institute of Physics of CNRS for supporting his stay at CNRS/Thales. We acknowledge synchrotron SOLEIL and HZB for provision of synchrotron radiation facilities.

### Author Contributions

Design and conception of the experiment: M.B. and M.N.G. Sample growth and characterization: M.N.G. STEM and EELS: G.S.S. and M.V. XAS, XMRS and XMCD measurements and data analysis: M.N.G., S.V., E.W., E.S., R.A., A.A., A.B., M.B. and J.S. Photoemission measurements and data analysis: M.N.G., J.R., J.P.R., J.S. and M.B. First-principles calculations: J.V. Article writing: M.B., M.N.G. and J.S. with inputs from all authors.

## Methods

***Growth of the heterostructures:*** The $LaAlO_3$ substrates were cleaned in acetone and propanol and annealed prior to deposition. All targets were pre-ablated in the growth conditions for 15 min. The GTO was then deposited by pulsed laser deposition at $PO_2 \approx 2.10^{-6}$ mbar and $T_{dep} \approx 700$°C. The growth was monitored by *in situ* RHEED and stopped after 7 u.c. Then the temperature was slowly decreased in the titanate growth pressure to the growth temperature for the nickelates (600-650°C). While the temperature was reduced, we monitored the evolution of the titanate with RHEED and no structural modification was observed. Right after deposition the samples were annealed in situ for 30 min at 500°C and 300 mbar of oxygen.

***TEM and EELS analysis:*** Cross sectional specimens for electron microscopy were prepared by conventional methods: grinding and Ar ion milling. STEM-EELS observations were carried out in a Nion UltraSTEM200 dedicated STEM equipped with a fifth order aberration corrector and a Gatan Enfinium spectrometer. On occasion, principal component analysis was used to remove random noise from EELS datasets. EELS maps have been produced both by integrating the signal under the edges of interest after background subtraction using a power law or by using a multiple linear least squares fit routine.

***X-ray absorption spectroscopy:*** The experiments were performed at the electron storage ring of the Helmholtz-Zentrum Berlin (HZB) by using the 70 kOe high-field end station located at the UE46-PGM1 beamline. Spectra were obtained across the Ti and Ni $L_{3,2}$ edges as well as across the O *K* and Gd and Nd $M_{5,4}$ edges. A magnetic field of 30 kOe was applied perpendicularly to the sample surface during measurements. Absorption experiments (XAS) were performed at normal incidence by means of total electron yield detection. The escaping depth of the secondary photoelectrons guarantees that the measured signal arises from the whole GTO and RNO thickness. Reflection experiments, *i.e.* XRMS and $A_{XRMS}$, were done in a Θ/2Θ geometry for which the sample was placed at an incidence angle of Θ =19° with respect to the incoming propagation direction. The XMCD data were taken at 5 K at normal incidence by means of total electron yield detection. XMCD was experimentally obtained as the difference between two absorption spectra measured with right ($\mu_+$) and left ($\mu_-$) circular polarized radiation whereas $A_{XRMS}$ is defined as the difference of the two reflection curves ($R_+$ and $R_-$) normalized by their sum. The Ni $L_{2,3}$ XAS spectra were background corrected at the pre-edge of the *L*-edge (far from the overlapping La $M_4$ signal) and then renormalized at the post edge. To suppress the La signal we corrected the XAS measured at the Ni *L*-edge in the bilayers using the XAS measured in the same conditions and same temperature from a reference LAO/GTO(7 u.c.) sample deposited on LAO. For the analysis of the Ni $L_3$ of the LNO reference sample, we subtracted a lorentzian function to the data as in Ref.[30]. The Ti *L*-edge signal was corrected by subtracting a linear background at the pre-edge and renormalizing it to the post-edge. For the sake of clarity, the spectra for reference samples were then shifted to align the second peak at the $L_3$ edge to be at the same energy. The O *K*-edge signal was also background corrected at the pre-edge and renormalized at its post-edge (550 eV).

***Hard X-ray Photo-emission spectroscopy:*** We have conducted hard x-ray photoemission spectroscopy (HXPS) experiments at GALAXIES beamline (Synchrotron SOLEIL, France)[44]. The photon energy was set at 2300 eV and the binding energy scale has been calibrated using the Fermi edge of the sample at 2295.00 eV. The overall resolution was better than 250 meV and all measurements have been done at room temperature. A Shirley background has been subtracted from every core-level spectrum. The angle *vs* depth photon field magnitude has been computed using YRXO software[45] taking into account the experimental geometry and the bilayer optical properties at 2300 eV for linearly polarized light. The modelling of the photon field as a function of the photon incidence angle is based on Chiam *et al.*'s methodology[46] using the depletion model of Tanaka *et al.*[47].

***First principles calculations:***

First principles calculations were performed using Density Functional Theory (DFT) calculations with the VASP package[48,49]. We used the PBEsol[50]+U framework as implemented by Lichtenstein's method[51]. The Hubbard U correction on Titanium and Nickel atoms was chosen to reproduce their bulk ground state properties. A value of $U_{Ti}$=3 eV previously yields correct ground state properties in strongly correlated $RTiO_3$ compounds[52]. The case of nickelates is more difficult to address and there is a debate on whether a small or large value of U is required to reproduce the ground state properties. We have chosen a $U_{Ni}$=2 eV, yielding the correct ground state properties regarding the complex antiferromagnetic (AFM) structure (T-type AFM or S-type AFM) and the band gap values on a large range of nickelates (R=Y, Gd, Sm, Nd and Pr). We emphasize here that $LaNiO_3$ was not considered in our DFT study, as it requires its own parameters to reproduce the bulk ground state (the ideal case would be $U_{Ni}$=0 eV as discussed in Ref. [40]). Geometry optimizations on a set of $(GdTiO_3)_n/(GdNiO_3)_n$ (n≤7) were performed until forces were lower than 0.01 eV/Å and energy was converged to $1.10^{-6}$ eV. All possible lattice distortions were considered in the calculations, including the breathing of the oxygen cage. The plane wave cut off was set to 500 eV and we used a k-point mesh of 5×5×1 for superlattices with $n \geq 3$ (5×5×3 for n=1). We worked at the collinear level and only a ferromagnetic solution was used in all our simulations of superlattices. PAW pseudo-potentials[53] were used in the calculations with the following electron configurations: $4s^2 3d^2$(Ti), $4s^2 3d^8$(Ni), $2s^2 2p^4$(O), $4s^2 4p^6 5s^2 4f^1$(Pr, Nd, Sm), $4p^6 5s^2 4f^1$(Gd) and $4s^2 4p^6 5s^2 4d^1$ (Y). We note that f electrons were not treated explicitly in our calculations and were included in the pseudo-potential. Results of charge transfer as a function of the rare earth in nickelates were obtained in a set of $(RTiO_3)_1/(RNiO_3)_1$ superlattices (R=Y, Gd, Y, Sm, Nd, Pr and Ce). This is justified by the fact that the charge transfer (i) appears to be mainly interfacial and (ii) is only controlled by the sole rare earth in the nickelate (cf. SI). It also avoids any built-in polarization altering the charge transfer.

## FIGURE CAPTIONS

***FIGURE 1. Growth and structural characterization.*** (a) RHEED monitoring during growth with intermediate RHEED images at each stage: LAO substrate before growth (left), GTO layer before deposition of the nickelate (middle), LNO layer at the end of growth (right). (b) Low magnification Z contrast image of the LNO/GTO heterostructure (top), and EELS maps from an area such as the one marked with a green rectangle (bottom). From bottom to top maps corresponding to the Ti $L_{2,3}$, La $M_{4,5}$, Ni $L_{2,3}$ and Gd $M_{4,5}$ edges are shown. (c) Atomic resolution Z contrast image of the same sample in the [110] orientation. The panel on the right shows the Ti $L_{2,3}$ (blue) and La $M_{4,5}$ (orange) EELS maps acquired on an area such as the one highlighted with a green rectangle.

***FIGURE 2. Interfacial charge transfer in $LaNiO_3$/$GdTiO_3$.*** (a) Evolution of the Ti and Ni valence across the interface from first-principles. (b) Variation of the Ti $2p_{3/2}$ binding energy in each Ti layer extracted from fitting the HXPS spectra shown in (d). (c) Variation of the O K-edge pre-peak area (large squares) and energy shift (small blue dots) resolved for each layer and extracted from the EELS spectra presented in (e). In (a), (b) and (c), the green and blue dotted lines are guides to the eye. (d) Experimental data (dots) and fit (thick solid lines) of the Ti $2p_{3/2}$ spectra measured at 0.2 deg and 10 deg (cyan and blue). The thin Gaussian lines show the individual simulated Ti $2p_{3/2}$ spectra for each Ti plane used to reproduce the experimental data. (e) EELS at the O K-edge at different positions in the GTO (lighter blue: closer to the interface). (f) Ratio of the intensity of peak a and peak b in (g). The width of the horizontal bars indicates which Ni planes contribute the most to the signal. The dotted line sketches the transfer of spectral

weight near the interface (see text for details). (g) Valence band HXPS at 0.2 (dark green) and 10 deg (light green) incidence angles. (h) Difference between the 10 deg and the 0.2 deg spectra. The arrow highlights the negative difference at the position of peak b.

***FIGURE 3. Tuning interfacial charge transfer by the rare earth in the nickelate.*** (a) XAS at the Ni $L_{2,3}$ edge for LNO/GTO (red), NNO/GTO (orange) and SNO/GTO (green). The black rectangle highlights the Ni $L_2$ edge. (b) Detailed view of the Ni $L_2$ experimental spectra (thick lines) in (from top to bottom) a LNO single film (violet), the three bilayers and a NiO reference (blue, from Ref. [24]). All spectra were fitted with two gaussian peaks (shown as thin lines). (c) Relative intensity of peak "a" in Fig. 3b plotted for the different samples (the dashed line is a B-spline passing through the data). (d) DFT calculations of amount of transferred electrons $\Delta N_e$ from the Ti ions across the interface in $(RTO)_1/(RNO)_1$ (for R= Ce, Pr, Nd, Sm, Gd, Y) relative to the value for R=Gd. (e) XAS at the Ti $L_{2,3}$ edge for a GTO film (purple), LNO/GTO (red), NNO/GTO (orange), SNO/GTO (green) and a STO crystal (blue). The black rectangle highlights the second peak of Ti $L_3$ edge used for the analysis. (f) Detailed view of the experimental XAS (symbols) at the second peak of the Ti $L_3$ edge for (from top to bottom) a GTO film, the three bilayers and a STO crystal. Except for STO, all spectra were fitted with two gaussian peaks (shown as thin lines ; the resulting spectra are shown as thick lines). The horizontal bar indicates the energy separation between the peaks. (g) Energy difference between the two gaussian components (the dashed line is a B-spline passing through the data). (h) DFT calculations of the variation of transferred electrons $\Delta N_e$ to the Ni ions across the interface in $(RTO)_1/(RNO)_1$ (for R= Ce, Pr, Nd, Sm, Gd, Y) relative to the value for R=Gd.

***FIGURE 4. Covalence* vs *ionicity.*** Experimental XAS signal (thick line) at the Ni $L_3$ edge for a LNO film and a LNO/GTO bilayer (a) and a SNO film and a SNO/GTO (b). All spectra were fitted with two peaks (thin lines) and shifted in energy to align the first component at 853.5 eV. The arrows indicate the position of the second peak. (c) Energy difference between the two components in (a), (b) and for a NNO film and a NNO/GTO bilayer analyzed similarly (not shown). (d) Covalence extracted from first-principles calculations of density of states in nickelate films strained on LAO (open symbols) and bulk (solid symbols) (e) O *K*-edge for reference thin films of SNO (violet) and GTO alone (blue) along with the same bilayers. In the bilayers data the integrated area for the Ni and Ti prepeaks are shaded in purple and blue, respectively. (f) Relative integrated area for the different bilayers. (g) FWHM of the Ni and Ti prepeaks. (h-j) Schematic representation of the interfacial charge transfer and the rehybridization processes in the covalent bond represented as two independent bands ($|3d^7\rangle$ and $|3d^8\underline{L}\rangle$) for decreasing

rare-earth size (from left to right). The upper panels represent the two parts of the heterostructure before contact and lower panels depict the resulting electronic state after charge transfer and rehybridization.

***FIGURE 5. Induced magnetic moment in the nickelates*** X-ray absorption, X-ray resonant magnetic scattering and X-ray magnetic circular dichroism spectra obtained at 8 K and 5 T in a NNO/GTO bilayer. XAS spectra were collected at the Ti (b) and Ni $L_{2,3}$ (d) edges and at the Gd (a) and Nd $M_{4,5}$ (c) edges. XMCD at the Gd $M_{4,5}$ and Ni $L_{2,3}$ edges are shown in panels (a) and (d), right axes. The asymmetry of the XMRS signal with respect to photon helicity measured at the Ti $L_{2,3}$ edge is also presented in panel (b), right axis.

***FIGURE 6. Role of covalence on magnetism*** (a-f) XRMS asymmetry measured at different fields ramping the field down from 3 T (darker graphs) and up from -3 T (lighter graphs) (g) Amplitude of peak b extracted from the XRMS asymmetry spectra as a function of the magnetic field (inset: same for peak a). (h) Highlight of the low field region. (i) Full Ni $L_{2,3}$ XMCD spectra acquired at 8 K for LNO/GTO (red), NNO/GTO (orange) and SNO/GTO (green). (j) Ni $L_2$ edge experimental spectra (symbols) for all three bilayers. The spectra were fitted by two gaussian components shown as thin lines, with the resulting fit displayed as a thick line. (k) Relative intensity of peak "a" in (j) plotted for the different bilayers.

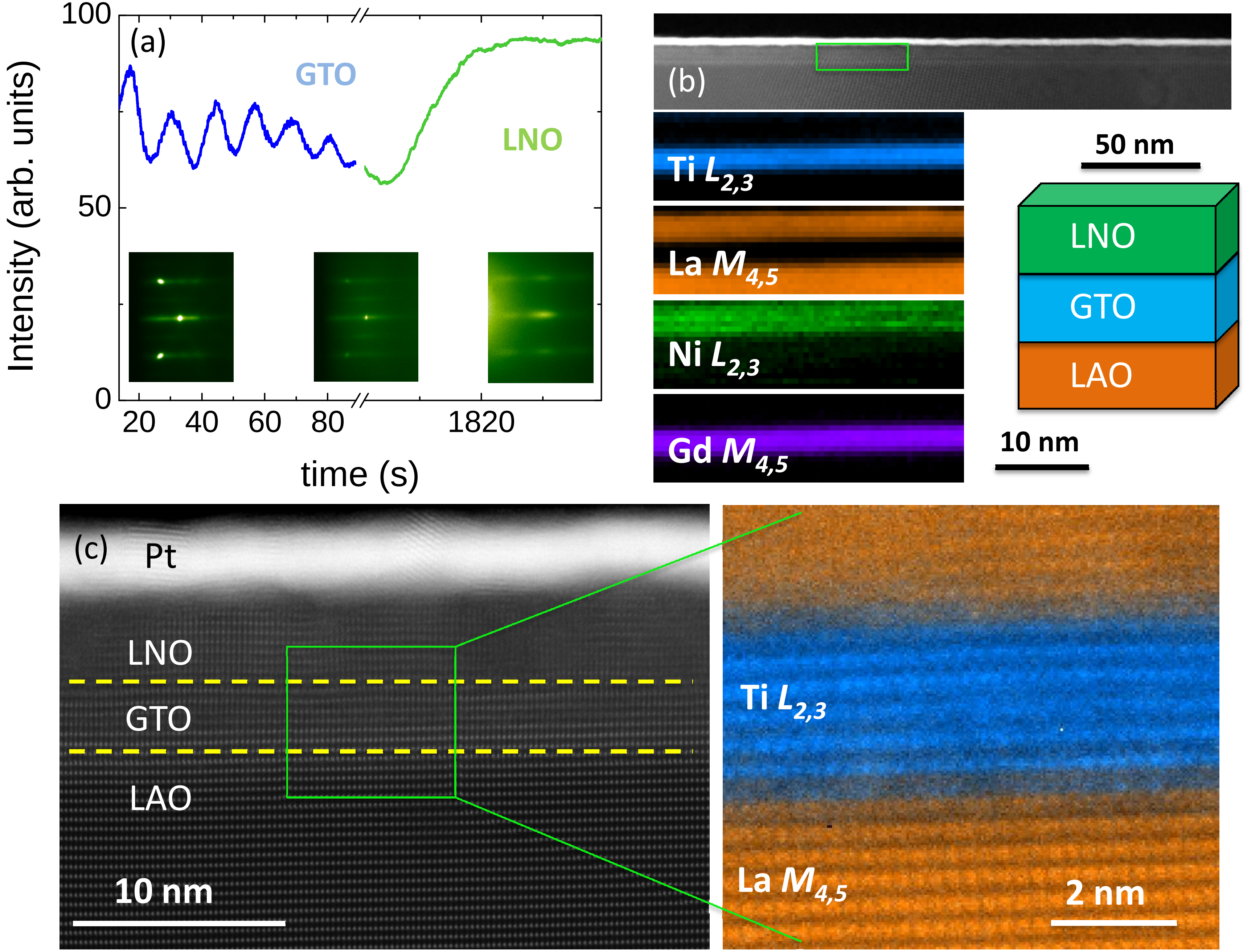

Fig.1

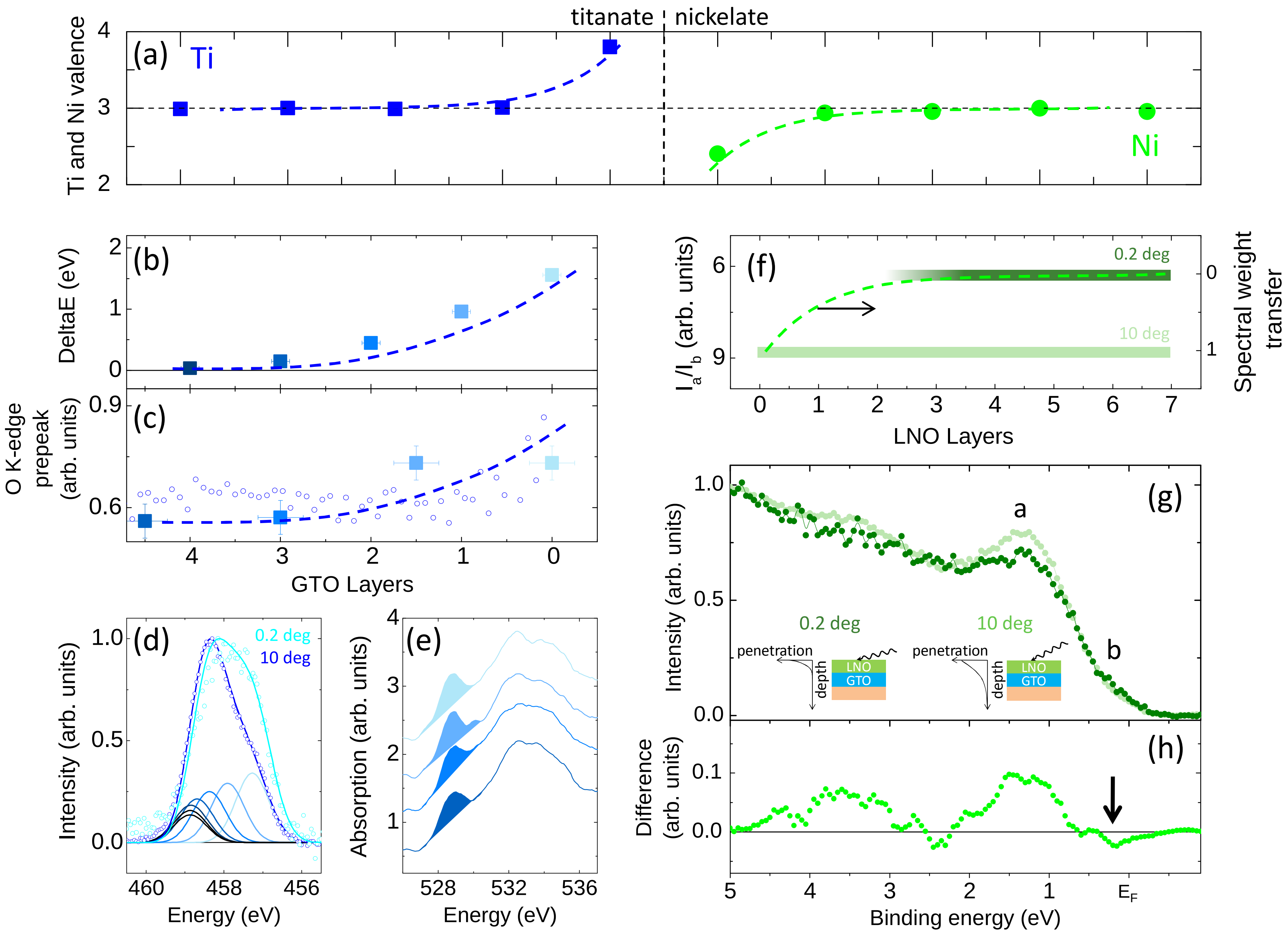

Fig.2

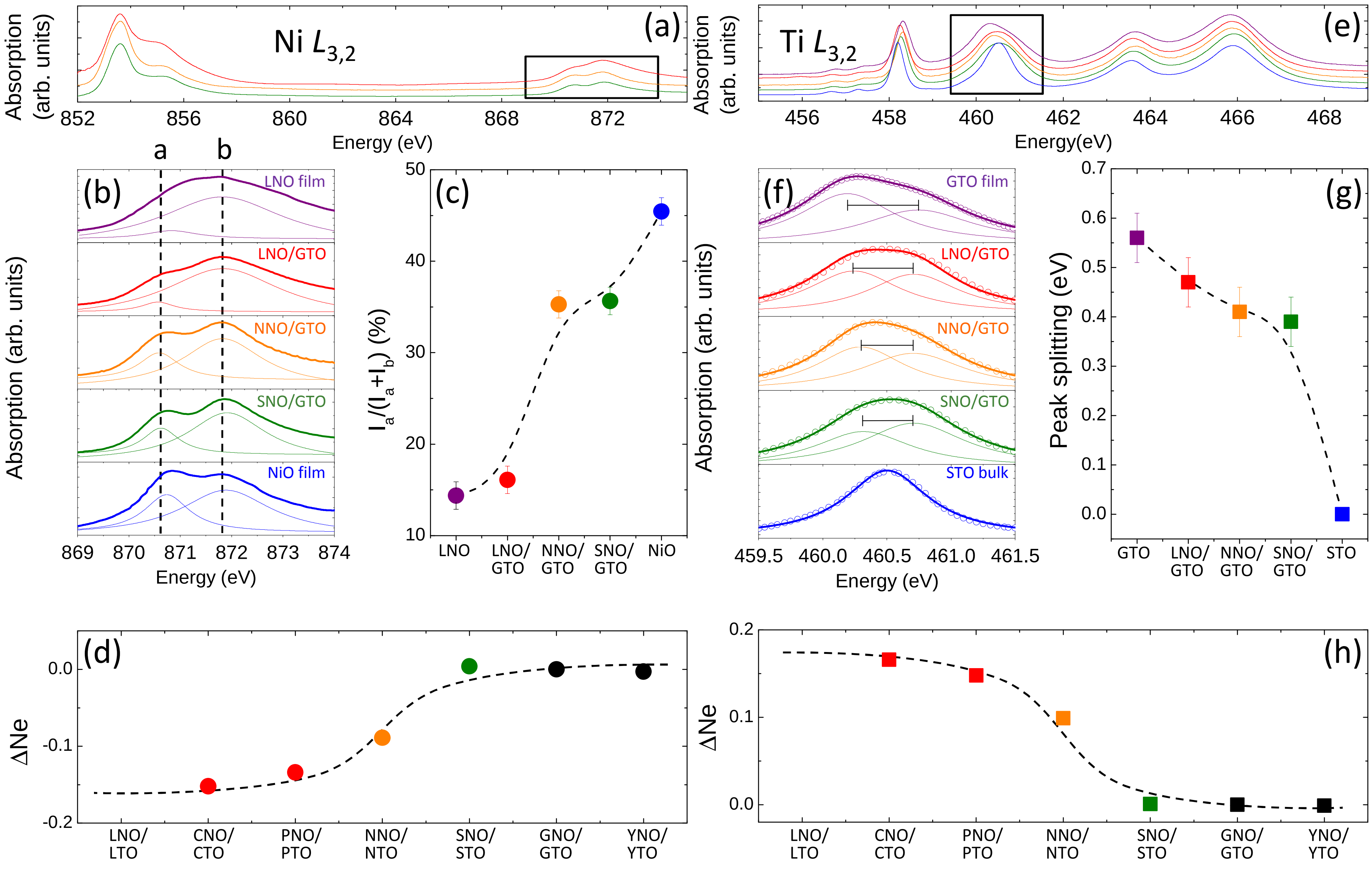

Fig.3

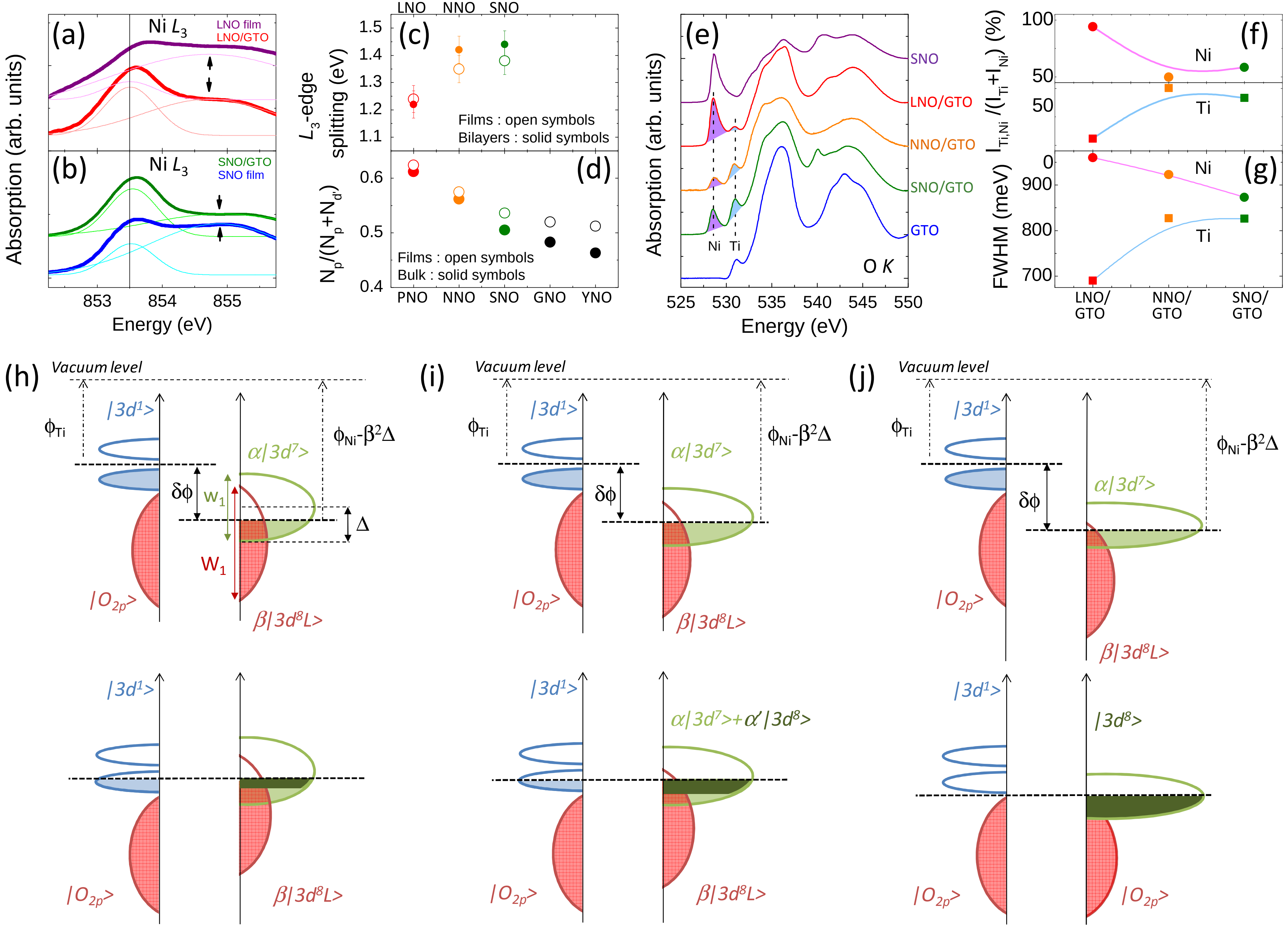

Fig.4

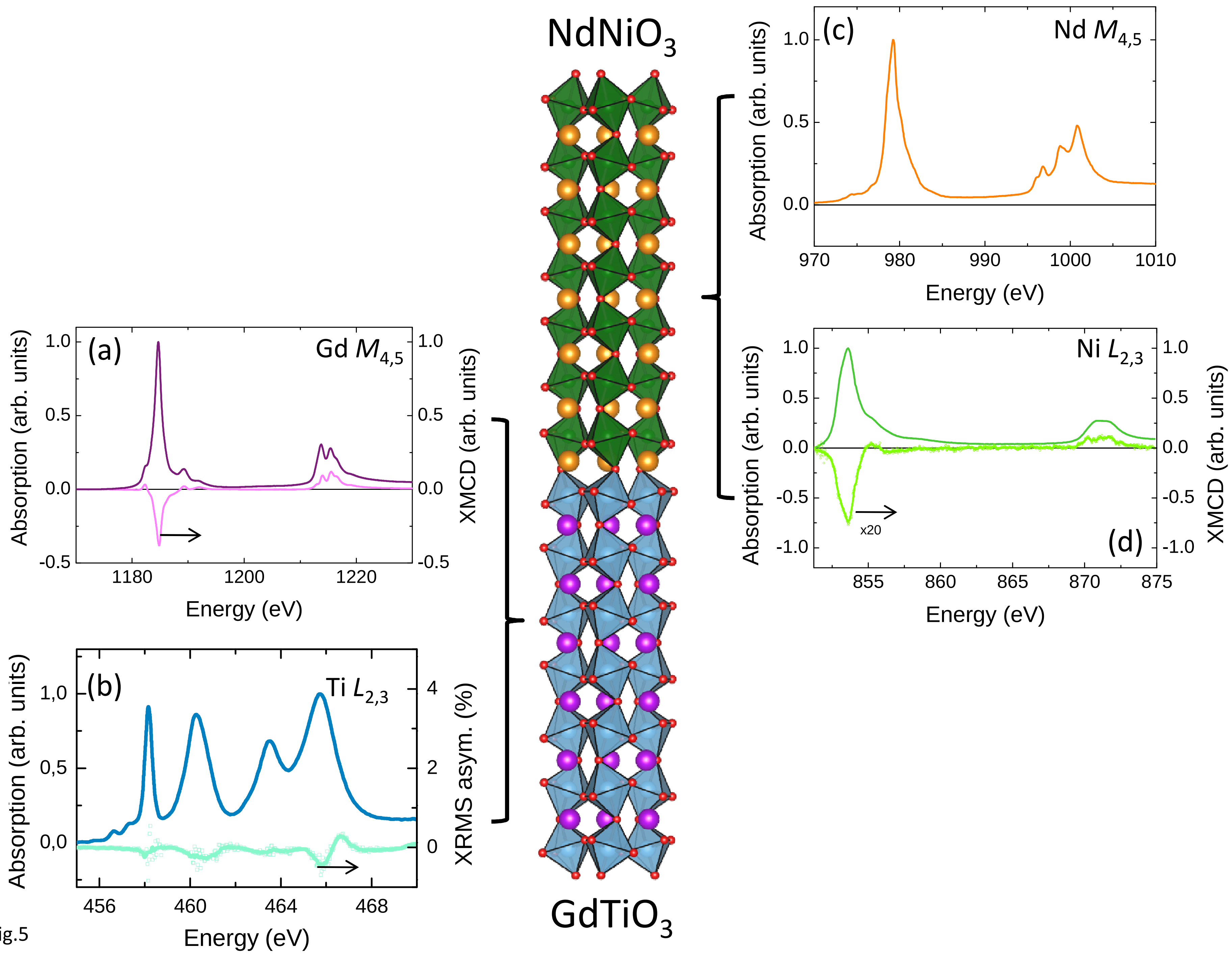

Fig.5

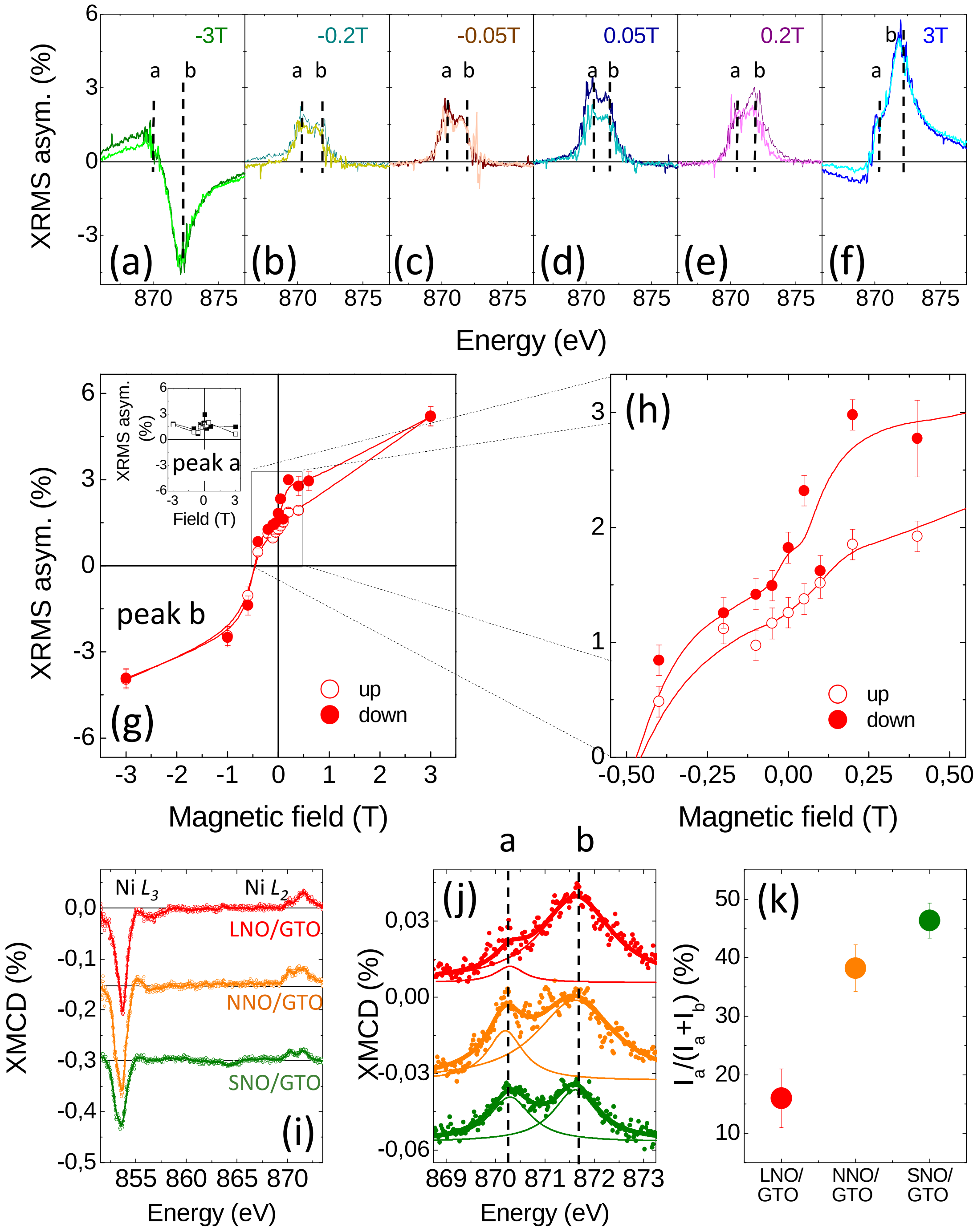

Fig.6